*Exploring the use of empirical measures to test the influence of research security policy on international collaboration in science*

# The Securitization of Science: A Systems Perspective on Policy and Measurement

**Conference on Geopolitics and International Collaboration in Science, Technology & Innovation, Manchester, June 1-3, 2026**

Caroline S. Wagner

**Table of Contents**

**The Securitization of Science: A Systems Perspective on Policy and Measurement**

Caroline S. Wagner[1]
*Academy Professor, John Glenn College of Public Affairs, The Ohio State University*

## Abstract

International scientific collaboration is governed primarily by researcher-level logic and network dynamics in which scientists seek partners at the frontier of their field with little consideration of national affiliation or geopolitical context. Research security policies that raise the friction costs of international collaboration are often assumed to operate against this logic, with a goal of producing selective withdrawal from sensitive fields in an effort to deny knowledge transfer to specific countries of concern. This paper tests that assumption against eight years of bibliometric evidence from China's scientific output (2018–2025) across 27 Scopus subject categories. China's internationally co-authored publication share declined universally across all fields after 2018--consistent with China's push for domestic capacity maturation--but the pattern of decline is inconsistent with the deterrence prediction. Fields with high dual-use proximity, including Computer Science, Materials Science, Physics and Astronomy, Chemical Engineering, and Engineering, showed the least contraction (3–6 percentage points), while fields with minimal dual-use salience--Psychology, Nursing, Dentistry, and Health Professions--showed the steepest declines (8–24 percentage points). An analysis of title and abstract language in China-affiliated computer science papers finds a simultaneous decline in security-sensitive vocabulary against a rising global baseline, consistent with adaptive participation at the paper level. Together the findings suggest that governments raised the friction level but did not redirect the network nor suppress collaboration in sensitive areas: researchers practiced adaptive participation, maintaining frontier collaborations while adjusting the form and language of their engagement. The pattern is precisely what invisible college theory, and the Ostrom knowledge commons framework predict.

[1] The Ohio State University, Columbus, Ohio. Wagner.911@osu.edu. ORCID: 0000-0002-1724-8489

## 1. Introduction: Securitization Policy Redraws the Language and Boundaries of Research Collaboration

A recent turn in science policy has been the addition of 'research security' terminology and policy to both the academic and political lexicon (Dao et al. 2024). The term 'research security' has gained increasing prominence in policy statements and law in the United States and Europe as governments react to the rise of China and the hyper-mobility of Chinese students and scholars seeking to work with foreign counterparts (Shih and Wagner 2024). James et al. (2025) argue that Europe is moving toward a more formalized "research security" regime, but remains fragmented, unevenly developed, and still searching for the right balance between openness and protection. A core challenge in policy documents is how to preserve the benefits of international scientific collaboration while reducing risks associated with geopolitical competition, espionage, foreign interference, intellectual property loss, and dual-use technology transfer. This paper proposes a framing of research security, its influence on science dynamics, and a test using bibliometric evidence.

The research security turn is best understood as activity at the interface between two coupled systems operating on different principles and at different speeds. On one side is the political support system. Modern public funding for science is sustained by a political justification, classically that private actors will not invest sufficiently in basic research, and it is therefore answerable to taxpayers and to internal political stakeholders whose perceptions can shift quickly. Public funders inside this system, the National Natural Science Foundation of China, the European Union and the national research ministries of Europe, and the National Science Foundation in the United States among them, supply resources and, as a condition of doing so, retain a claim over the terms on which those resources are distributed and applied. This is the contemporary form of a much older relationship between science and its practitioners, once organized through personal patronage and now organized through public accountability; the funder differs in everything but the basic claim. On the other side is the research community. Researchers form collaborative ties from the bottom up, selecting partners on the basis of reputation, complementary expertise, and the prospect of scientific reward. The structure they produce, what Crane and others have called the invisible college, grows through preferential attachment and accumulates over long time horizons. It does not respond to the same signals as the political support system, and it does not respond at the same speed.

The two systems are coupled through public money and support. The current research security turn is what happens when the political support system attempts to redraw the language and boundaries of legitimate connection, extending its claim from military research into basic research, while the research community continues to operate by its own selection criteria. The impact can be imperfectly measured by examining changes to the outputs of the knowledge system using published works.

Posed this way, the question is not whether research security policy works. It is structural: where does the political support system's redrawn boundary register in the observable record, and where does the research community continue to operate on its own terms? The contribution of this paper is therefore twofold. It offers a theoretical reframe, situating the research security turn at the interface between two coupled systems with different selection criteria, language, and time constants, rather than as an external shock to a what might be called a "stateless commons." And it offers a diagnostic, using the bibliometric record to locate that interface. The paper does not offer a causal estimate of policy effect, and Section 6 is explicit about why the available data cannot support one.

The paper proceeds as follows. Section 2 develops the theoretical framework, drawing on the political support system, the self-organizing structure of the global collaboration, and the speed and selection differences between the two. Section 3 describes the data and methods. Section 4 presents the empirical findings: a system effect that holds across the global network, a field-level pattern of defense-proximity, and a test of linguistic adaptation in word usage at the paper level in computer science. Section 5 interprets the pattern. Section 6 discusses limitations of this approach.

## 2. Securitization as boundary-redrawing, and a derived expectation

Securitization theory has been variously defined as the process by which political actors reframe an issue as an existential or critical threat requiring extraordinary measures (Balzacq, 2011; Eroukhmanoff 2020) and as the language applied to specific actions (Buzan et al., 1998). Applied to scientific collaboration, securitization recasts coauthorship, mobility, and funding ties as vectors of national exposure. Within the present framework, this is the public-accountability exercising agenda-setting power: redrawing the boundary of legitimate connection and reframing the language in which connection is understood. The framework does not treat securitization as a strawman to be refuted. It treats it as the mechanism. The open question is how far the mechanism reaches and how it affects the dynamics of collaboration in science.

The self-organizing logic of the knowledge system, operating as a network, generates an expectation about the dynamics that can be measured. If the network self-organizes around scientific value and the value of reciprocal exchange, then external political pressure will not dissolve ties uniformly. It will dissolve the ties where scientific value is lowest relative to the friction, which are the substitutable, peripheral ties sustained by convenience or institutional momentum. It will leave intact the high-value, complementary ties whose scientific cost of severance exceeds the patron-imposed cost of maintenance. The expectation, then, is that patron pressure registers at the periphery and has less influence at the core. This is not a claim that the invisible college as an institution is invulnerable. It is a narrower, testable proposition about the observable co-authorship record, and it is what the self-organizing dynamics of the college would lead one to expect.

## 3. Data and Methods

The analysis rests on two bibliometric sources. Absolute publication counts are not comparable across countries of different size, and changes in those counts can reflect system-wide growth rather than any bilateral dynamic. The analysis therefore foregrounds relative measures: collaboration share, field-relative change, and within-field comparison. The interpretive weight falls on shares and on patterns of divergence across fields, not on raw counts.

Data on China's output, citation impact, and international collaboration share are drawn from Scopus bibliometric data provided by Elsevier, covering up to 24 subject categories, with annual observations from 2018 to 2022. Publication counts are disaggregated by collaboration type, internationally co-authored versus domestic-only, which allows field-weighted citation impact (FWCI) to be computed separately for each. Two limitations are built into the analysis: FWCI values for 2023 to 2025 are treated with caution because of citation lag, so 2022 is used as the primary endpoint for citation impact. To add analysis, each subject category for China's outputs was assigned to a high, medium, or low dual-use tier based on explicit inclusion in research security targeting, documented dual-use concerns, and the directness of application pathways to security-relevant technologies; the classification is coarse and intended as a high-level test of field-selectivity; a fine-grained decomposition should be conducted in the future.

The design supports both descriptive and pattern-level observations. It can establish whether collaboration shares moved, in which direction, and whether the field-level pattern of movement runs with or against the pattern a research security protocol would predict. The analysis does not include policy variables or counterfactuals at this time. The government's most direct instruments act on inputs: on money, through funding conditions, and on people, through mobility and visa gatekeeping. Co-authorship is an output, observed one step downstream of the instrument and after a publication lag. The co-authorship record is thus an indirect trace of patron action, and as Section 5 argues, that indirectness is part of the explanation for why the boundary looks the way it does.

## 4. Results: The System Effect: Universal Share Decline Without Withdrawal

China's total scientific publication output in Scopus grew from approximately 604,000 publications in 2018 to 1.31 million in 2025, a 54 percent increase over seven years. Domestic-only output drove the majority of this growth, expanding from roughly 464,000 to 1.08 million, shown in Figure 1. By 2025, domestic-only publications accounted for approximately 82 percent of China's total output, up from 77 percent in 2018.

Citation impact tracked this expansion and shown in Figure 2. China's FWCI for internationally collaborative publications remained consistently above the world average throughout the study period, between roughly 1.72 and 1.86 from 2018 to 2022. More consequential is the trajectory of domestic-only FWCI, which rose from 0.849 in 2018 and crossed the world-average threshold around 2024 to 2025, reaching approximately 1.01, demonstrating a marked improvement in China's scholarship.

The gap between internationally collaborative FWCI and domestic FWCI did not close, which indicates that international collaboration continues to produce higher-impact work. China has not outgrown the benefit of collaboration; it has developed enough domestic capacity that non-collaborative work is no longer below the global mean.

*Figure 1 China's Total Publications and Share of Publications Coauthored at the International Level Data: Scopus*

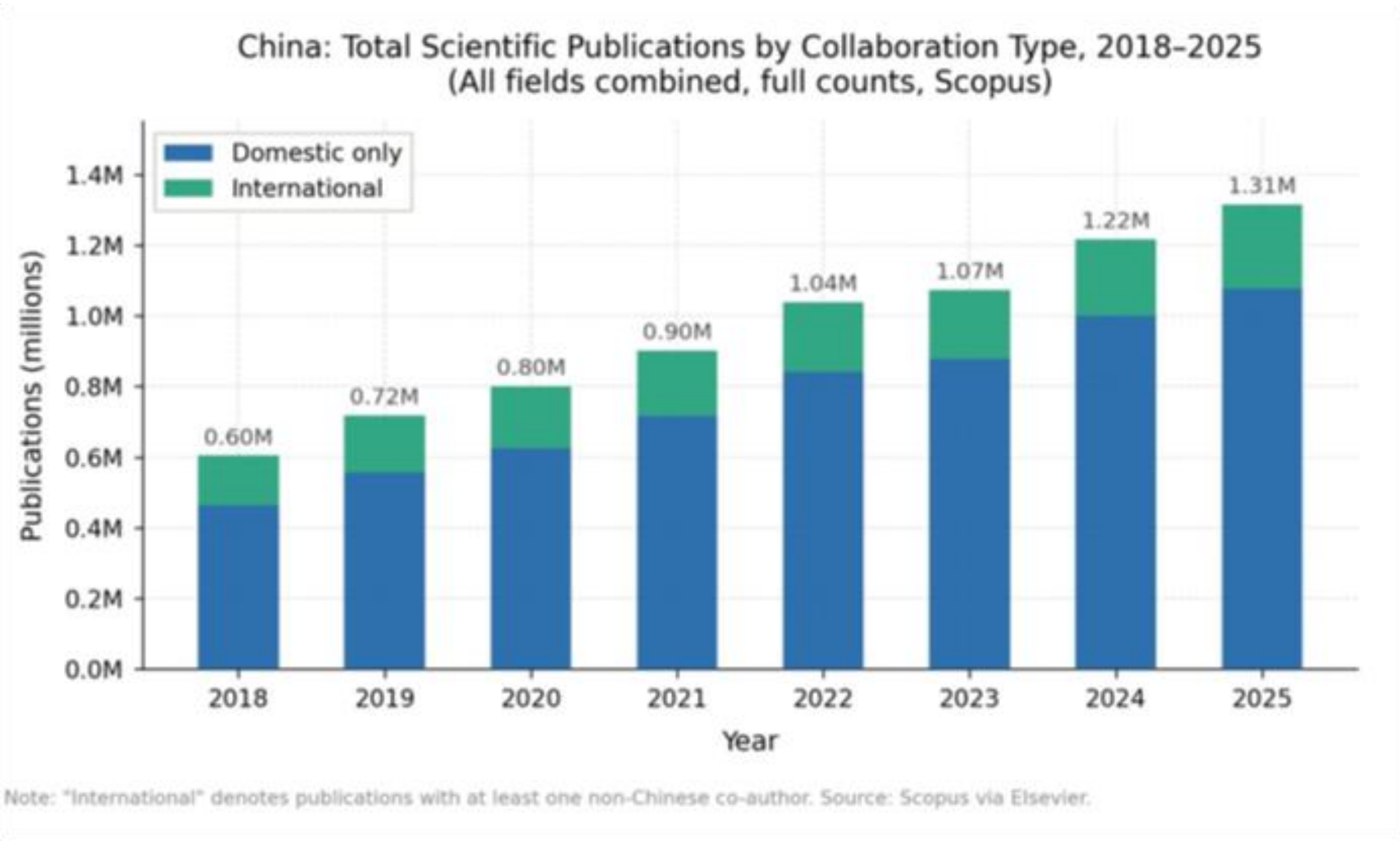

*Figure 2 Field-Weighted Citation Impact, Domestic and International Collaboration Data: Scopus*

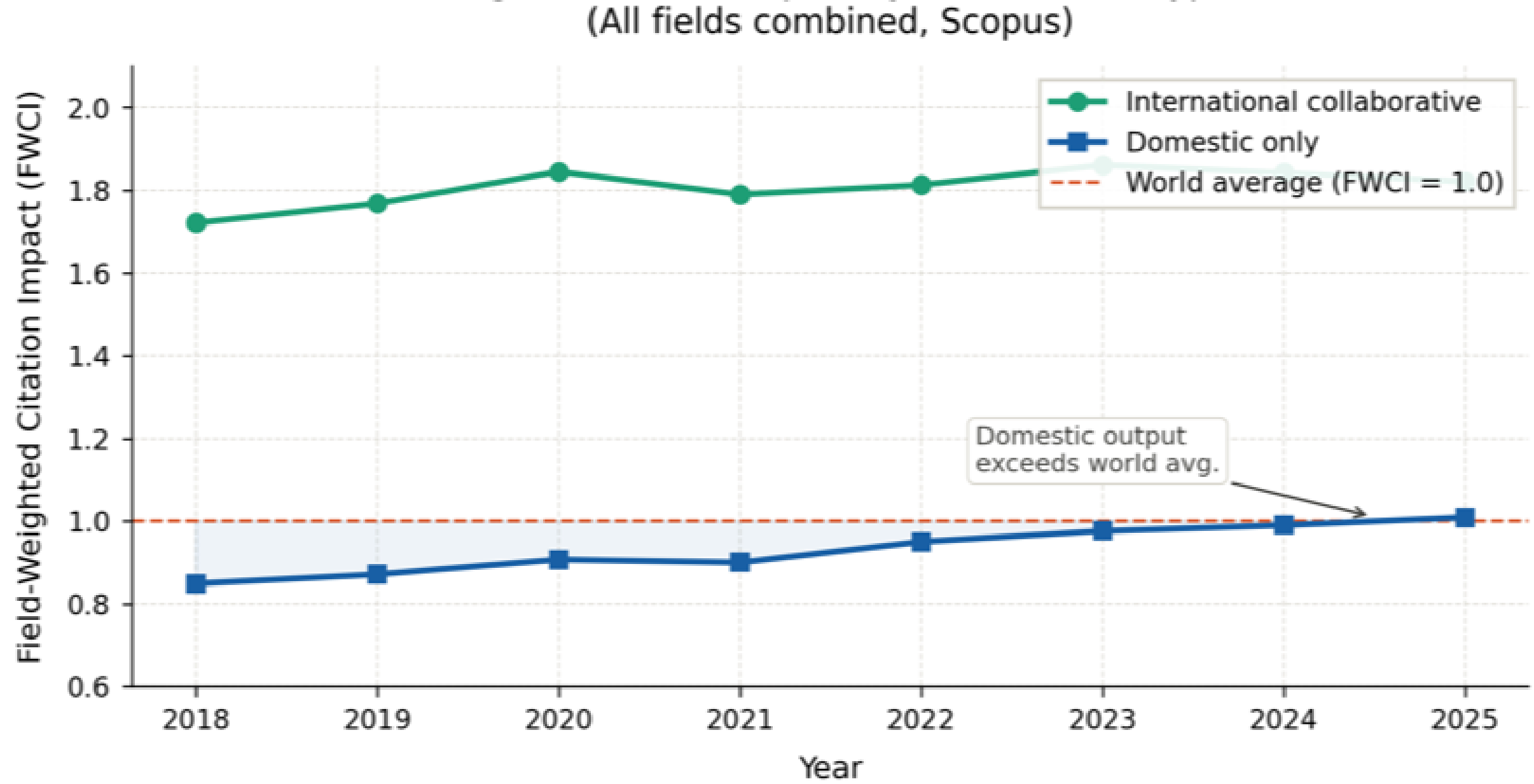

Note: FWCI > 1.0 indicates above-world-average citation impact after controlling for field and publication year. Source: Scopus via Elsevier.

4.2 The universal decline in collaboration share

Across all 20 analyzable Scopus subject categories, shown in Figure 3, China's share of internationally co-authored publications declined between 2018 and 2022. The magnitude ranged from a 2.7 percentage point decline in Engineering to declines of 11.5 points in Neuroscience and 17.3 points in Psychology. The universality of this decline is the analytically important fact. A decline that appears in every field, regardless of that field's dual-use profile or its exposure to explicit policy targeting, is difficult to reconcile with an account in which post-2018 changes are driven primarily by field-specific policy effects. It is far more consistent with a system-wide dynamic. As the volume figures show, domestic output grew faster than internationally collaborative output. A declining share under these conditions does not mean fewer international collaborations. It means the domestic base grew more quickly than the collaborative base.

This is the system effect. It is the predicted signature of a maturing participant in a self-organizing network, and it is the first piece of evidence about the limits of policy reach. The most visible aggregate change in China's collaboration profile, the falling international share, is not in the first instance a story about policy. It can also be attributed to endogenous capacity growth within China. Any account that reads the falling share as withdrawal under policy pressure has misread an arithmetic consequence of domestic maturation as a behavioral response to securitization.

*Figure 3 Drop in China's International Collaboration by Field, 2018-2025 Data: Scopus*

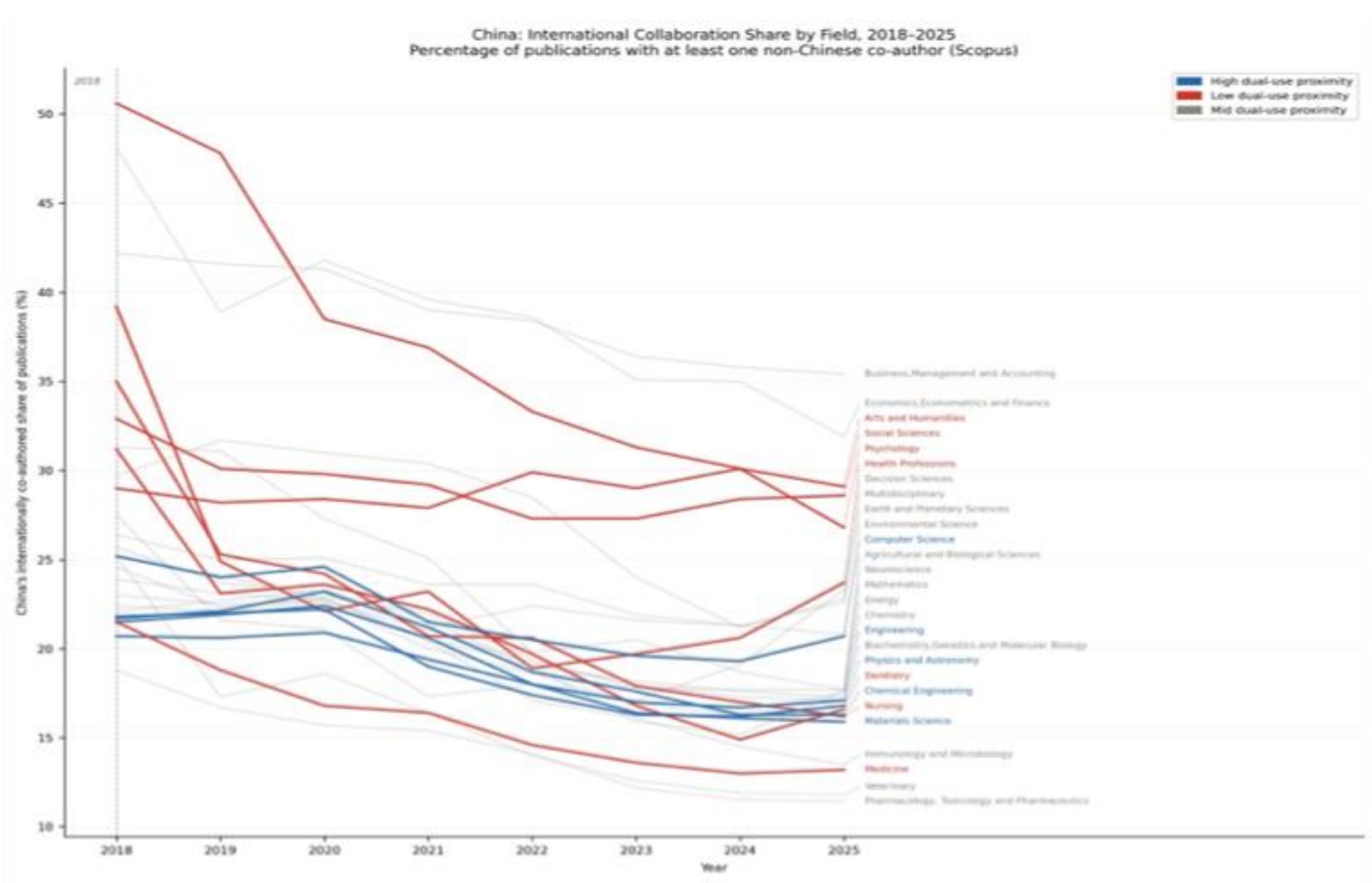

The drop in numbers of international collaborations is not uniform across fields, nor is it directly related to citation impact by field. Figure 4 shows the same data as Figure 3, but it includes the citation impact by field. Bubbles on either end of the curve show the relative size of contribution and the position of the field weighted citation impact for that year. Three health-related fields were above the global FWCI in 2018 (dentistry, veterinary, and health professions). By 2022, all fields grew in FWCI over to fall above the global FWCI by 2022. The curve shows the extent of growth within each field. Medicine shows the most rapid growth. The declining international share in health fields is driven primarily by domestic capacity growth rather than policy-induced deterrence or a drop in interest in working with China.

*Figure 4 Change in China's International Collaboration by Field and Impact   Data: Scopus*

4.3 The dual-use-proximity pattern

Setting the field-level changes against a dual-use classification list produces the central diagnostic finding. The fields most directly associated with dual-use research (and most explicitly targeted by US research security policy as proxy) did not show the steepest retreats from international engagement. Figure 5 shows the results by field as assessed against the dual-use criteria as a change in percentage of internationally coauthored works. China's Computer Science, Materials Science, Physics and Astronomy, Chemical Engineering, and Engineering expanded their internationally co-authored publication share over the longer 2008 to 2022 window. The steepest declines fell instead on clinical, health, and social fields with minimal dual-use profiles, including Dentistry, Health Professions, and Psychology. The gradient runs opposite to the direction a simple deterrence account would predict. The field-selective pattern is neither a paradox nor a refutation. It is precisely what the invisible college dynamic predicts under conditions of externally imposed friction. Governments raised the friction level--through disclosure requirements, funding restrictions, and the reputational costs of China-linked research activity. They did not direct the network. Researchers responded through adaptive participation: some shifted with whom they worked, routing toward partners at the frontier where the scientific value of the collaboration justified absorbing the friction; some reframed their titles and abstracts to describe their work in less politically exposed terms (see below); all continued. The invisible college logic--as in, seek the most valuable partners, work at the frontier, maintain the ties that advance the science--did not stop operating. Adaptive participation is not a departure from that logic. It is that logic applied under new constraints. Psychology, Nursing, Dentistry, Health Professions, and Neuroscience contracted most steeply--declining 8 to 24 percentage points--because those ties were already peripheral in invisible college terms, sustained by training relationships and institutional momentum rather than frontier scientific interdependence, and not sufficiently valuable to motivate adaptation.

Chemistry, Materials Science, Chemical Engineering, Energy, Physics and Astronomy, Computer Science, and Engineering, by contrast, showed the greatest resilience, declining only 3 to 6 percentage points. These are the fields where the frontier moves fastest, where Chinese researchers have developed the deepest complementary capabilities, and where the invisible college logic of scientific value was strong enough to absorb the friction and sustain the collaboration. What the pattern shows is not that the patron failed, but where the patron's agenda-setting power reached its limit.

*Figure 5 Field Selective Engagement Accounting for Defense Proximity Data: Scopus*

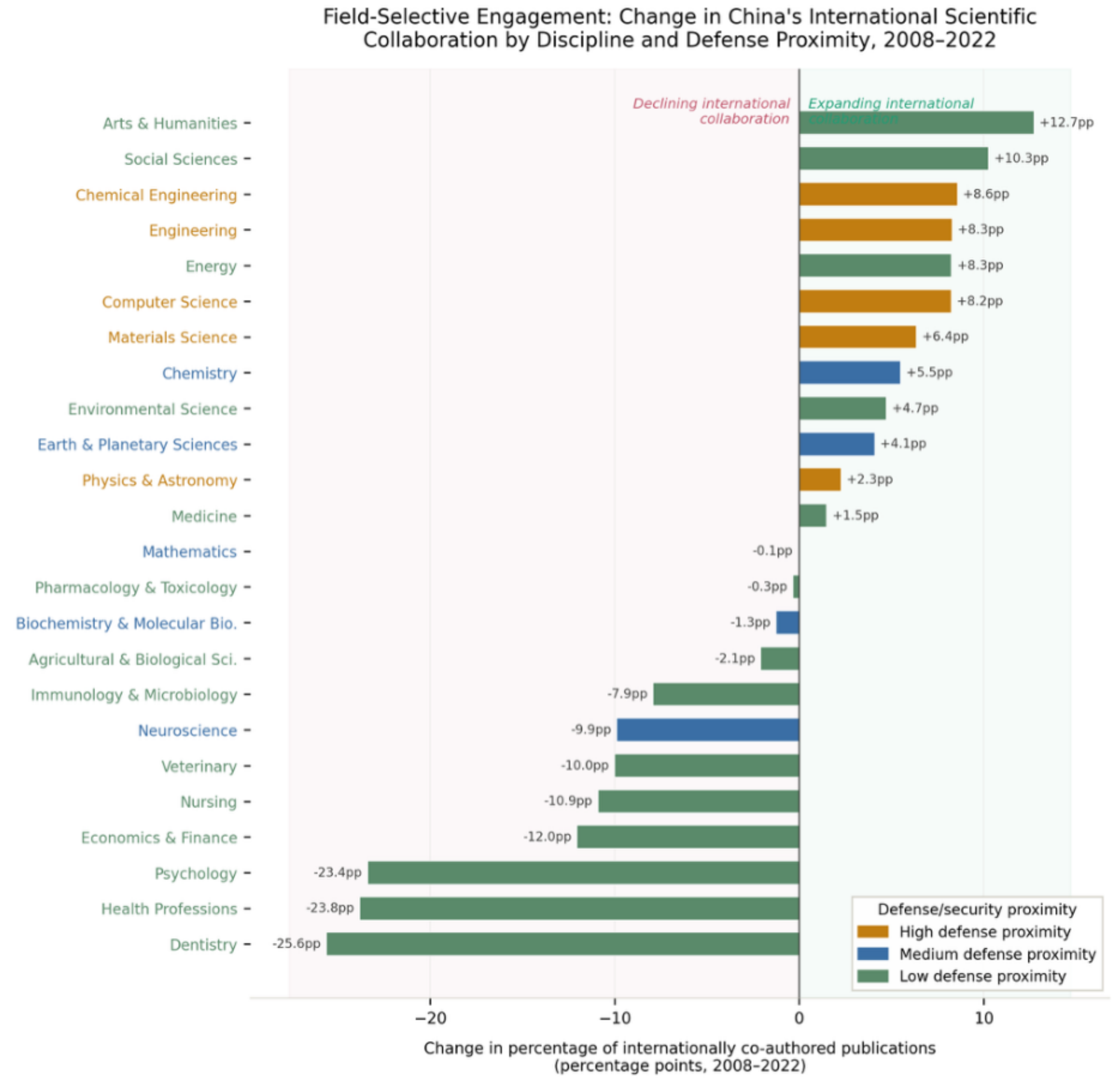

Note: Values show change in the share of internationally co-authored publications (percentage points). Defense proximity tiers reflect documented dual-use potential based on U.S. research security policy literature. Source: Scopus via Elsevier.

4.4 Language adaptation in computer science publications

The bibliometric evidence in the preceding sections establishes erosion and resilience in the co-authorship record. A complementary question is whether the boundary also registered in language--in the words researchers chose to describe their work. The Copenhagen School's foundational account of securitization holds that security is enacted through speech acts: an issue becomes a security issue when it is successfully framed as an existential threat, and that framing is produced and sustained through language (Buzan, Wæver and de Wilde, 1998). If securitization operates through language at the policy level--recasting co-authorship as national exposure, reframing scientific mobility as a security risk--then adaptation to securitization should also be visible in language at the researcher level. Researchers might be expected to learn the new terms acceptable for the purposes of political support, whether consciously or not, by adjusting their own language to reduce their exposure to it.

To test this, we examined title and abstract language in computer science publications before and after the 2018 policy inflection, comparing China-affiliated CS papers against the global CS baseline. Computer science was selected because it is the field with the highest volume of China-involved international collaboration in the dataset, the field most explicitly targeted by US research security instruments, and the field whose application spectrum spans most directly from civilian to dual-use. The analysis draws on a sample of approximately 4,000 China-affiliated CS papers from 2014–2017 (pre-period) and 5,000 from 2019–2023 (post-period), with the global CS baseline comprising 6,400 and 8,000 papers respectively. Abstract text was retrieved from the OpenAlex API. Data are preliminary and the sample is a fraction of the full population; findings should be treated as indicative rather than definitive.

The first finding is essential context for everything that follows. Across the global CS literature, dual-use terminology did not decline after 2018--it increased substantially. The average change in a pre-defined set of dual-use terms across the global CS baseline was +86% from the pre- to post-period. Specific terms grew sharply: drone (+147%), autonomous (+79%), LiDAR (+192%), anomaly detection (+181%), radar (+65%). This reflects the field-wide expansion of AI-driven autonomous systems, remote sensing, and adversarial machine learning that characterized computer science in this period. The global trajectory moved toward more dual-use language, not less.

This rising global baseline is the essential control. Any claim that securitization produced language change must show that China-affiliated papers diverged from this trend--not merely that some terms declined in absolute frequency, which could reflect field maturation or topic substitution unrelated to securitization.

Against the rising global baseline, China-affiliated CS papers show a pattern of divergence consistent with strategic linguistic adaptation under securitization pressure. The overall average change in dual-use terminology in China-affiliated papers was +122%, higher than the global baseline of +86%--driven by increases in drone (+629%), UAV (+396%), and adversarial (+435%) terminology, which are fields where Chinese research expanded rapidly. But within this overall increase, a cluster of terms declined sharply and specifically. Facial recognition fell to zero--a −100% decline--in China-affiliated papers, against a −53% decline in the global baseline. Spoofing declined −89% (versus −27% globally). Malware declined −83%, authentication −72%, cryptography −63%. Jamming--a term almost absent from civilian CS--declined −51% in China-affiliated papers while increasing +382% globally.

These are not terms that declined globally and declined more sharply in China-affiliated papers--which might reflect a general field shift. Several of these terms--jamming, cryptography, authentication, spoofing--either increased or held steady in the global baseline while declining specifically in China-affiliated work. The divergence from the global trend is the signal.

The mirror of the dual-use decline is a pronounced shift toward civilian-application language in China-affiliated papers. The average increase in civilian terms in China-affiliated papers was +159%, compared to +63% in the global baseline. Agriculture increased +1,580% in China-affiliated papers versus +289% globally. Education increased +511% versus +242%. Sustainability increased +363% versus +76%. Natural language processing rose +199% versus +135%. Drug discovery rose +149%. See Appendix table. Whether the substance of research changed is a subject for future research.

The vocabulary analysis corroborates this directional shift. Words specifically increasing in China-affiliated papers but not in the global baseline include: intelligent, digital, green, edge, technologies--terms associated with civilian, commercially acceptable, and geopolitically neutral application domains. Words specifically declining in China-affiliated papers but not globally include: nonlinear, fuzzy, distributed, fault, simulation, graphene--more technical engineering vocabulary with dual-use associations in several of its applications.

## 5. Interpretation: Agenda-Setting Power Meets a Self-Organizing College

This analysis set out to use empirical measures to test whether research security pressures were influencing China's international collaborative patterns, especially in sensitive or critical research areas, and to ask whether the measures are helpful. Clearly, government policies raised the transaction costs of collaboration, imposed disclosure requirements, restricted mobility, and reframed co-authorship as a security exposure, and in the high-dual-use fields it applied that pressure with particular intensity through targeted scrutiny. One would expect collaborations to decline, which they did. However, one would expect collaborations to decline in the most sensitive fields, which they did not.

Moreover, the decline evidenced in China's international collaboration is more a factor of China's growth in domestic output than it is in international collaboration, which continued to grow at a very low rate. Fall-off in collaborations are seen in Psychology, Nursing, Dentistry, Health Professions, and Neuroscience, where Chinese policy has encouraged much more domestic research. Collaborations do not fall off in Computer Science, Materials Science, Physics and Astronomy, Chemical Engineering, and Engineering, which include research closer to dual-use applications than the fields that dropped off.

The pattern of word usage supports an adaptive participation hypothesis in which actors redirect language while retaining international linkages. It is consistent with a form of desecuritization performed by researchers at the paper level: a linguistic repositioning of their work outside the securitized zone around which policy has drawn certain boundaries.

The evidence indicates that agenda-setting power can raise friction across the board but cannot, on its own, redirect epistemic choice at the high-value core. The instruments are therefore mismatched to the stated objective in a specific way. They are most effective at dissolving the peripheral, lower-value ties that are of least security concern, and least effective at the high-value ties that are the primary objects of concern.

**6. Limitations: What the Data Cannot Show**

This approach presents limitations to be considered as we seek to measure the dynamics of change infused by research security.

First, the analysis taps only published scientific literature. Research that has been classified, restricted, or moved into non-publication channels is invisible to bibliometric analysis. If genuine decoupling is occurring in those channels, this study cannot see it.

Second, the co-authorship record cannot capture collaborations that were never formed. A chilling effect that deters a partnership before it produces a paper leaves no trace in the data. The observable record is therefore a lower bound on patron effect, not a measure of it. The overall decline in international collaboration counts is documented in the broader literature, but the extent to which that decline is attributable specifically to securitization policies, as opposed to domestic capacity growth or other factors, cannot be determined from co-authorship data alone.

Third, there is no identification strategy. The study documents that collaboration patterns changed and that the field-level pattern of change runs against the deterrence prediction. It does not, and cannot, attribute any specific change to any specific policy instrument. Publication counts and shares move for many reasons: endogenous capacity growth, field maturation, the pandemic, and broader geopolitical conditions among them. Policy-attributable change cannot be isolated from these without an identification design the present data do not support.

Fourth, the endogenous and exogenous components of the post-2018 reorganization cannot be cleanly separated. Some portion of the change reflects the network self-organizing around China's growing domestic capacity, a process already underway before the policy escalation. (See Wagner & Cai, 2026.) Another portion reflects the policy friction itself. The direction of the evidence, with high-value ties persisting and peripheral ties dissolving, is consistent with the initial hypothesis, but the data cannot deliver a clean decomposition of influences.

Fifth, the claim is bounded to the published co-authorship record and to field-level aggregation through roughly 2024. It is not a claim about the resilience of the invisible college as a social institution. The college may be robust, fragile, or changing in ways that field-level co-authorship counts cannot register. What the data support is the narrower statement: in the observable record, patron boundary-setting registered at the periphery and slipped at the core.

## Appendix: Data and Methods Notes

*A.1 Scopus anomalous 2017 values*

Seven fields showed values in the 2017 snapshot inconsistent with adjacent years and likely reflecting data extraction issues rather than real trends: Energy, Materials Science, Physics and Astronomy, Veterinary, Health Professions, Pharmacology and Toxicology, and Economics, Econometrics and Finance. A verification and annual-data request has been submitted to Elsevier. Affected fields are excluded from series where their inclusion would distort the trend.

*A.2 Defense-proximity classification*

High tier (Computer Science, Engineering, Materials Science, Chemical Engineering, Physics and Astronomy): explicit inclusion in US Export Administration Regulations and the National Security Presidential Memorandum on critical and emerging technologies; documented dual-use literature; direct application pathways to military or intelligence capability. Low tier (Medicine, Nursing, Dentistry, Health Professions, Social Sciences, Arts and Humanities): absence of all three criteria. Medium tier: remaining fields. The classification is coarse at the field level; topic-level classification is identified as a priority for subsequent analysis.

*A.3 Word Frequency and Changes Related to Security or Dual-Use Fields*

**A. 3 Table: Language Shift in Computer Science Publications, 2014–2017 vs 2019–2023**

*Percentage change in term frequency (per 1,000 words) in title and abstract text. Global CS baseline = all CS papers worldwide. China-affiliated CS = papers with at least one Chinese institutional address. Red = decline ≥30%; Green = increase ≥30%. Divergence = China change minus global change.*

| Term / Application domain | Global CS 2014–2017 vs 2019–2023 | China-affiliated CS 2014–2017 vs 2019–2023 | Divergence from global trend |
|---|---|---|---|
| ***Panel A: Dual-use / security-sensitive terms--Declining in China-affiliated papers*** | | | |
| Facial recognition | **-52.9%** | **-100.0%** | **-47.1%** |
| Spoofing | -26.7% | **-89.3%** | **-62.6%** |
| Malware | -11.3% | **-82.7%** | **-71.4%** |
| Authentication | -13.8% | **-72.1%** | **-58.3%** |
| Cryptography | -6.2% | **-63.1%** | **-56.9%** |
| Jamming | **+382.4%** | **-50.6%** | **-433.0%** |
| Surveillance | +16.9% | **-45.4%** | **-62.3%** |
| Tracking | -19.0% | **-35.0%** | -16.0% |
| Face recognition | **-31.9%** | **-31.3%** | +0.6% |

| Term / Application domain | Global CS 2014–2017 vs 2019–2023 | China-affiliated CS 2014–2017 vs 2019–2023 | Divergence from global trend |
|---|---|---|---|
| Swarm | **+65.6%** | **-31.1%** | **-96.7%** |
| Cyber | **+88.9%** | -18.2% | **-107.1%** |
| ***Panel A continued: Dual-use terms rising in both groups (field-wide AI expansion)*** | | | |
| LiDAR | **+192.4%** | **+221.9%** | +29.5% |
| Autonomous | **+79.1%** | **+246.0%** | **+166.9%** |
| UAV | +17.5% | **+396.1%** | **+378.6%** |
| Drone | **+146.8%** | **+628.7%** | **+481.9%** |
| Adversarial | **+114.0%** | **+434.6%** | **+320.6%** |
| Generative adversarial | **+138.9%** | **+216.8%** | **+77.9%** |
| Navigation | **+47.8%** | **+150.6%** | **+102.8%** |
| Anomaly detection | **+181.1%** | **+45.0%** | **-136.1%** |
| ***Panel B: Civilian / neutral-application terms--Accelerating in China-affiliated papers*** | | | |
| Agriculture | **+288.9%** | **+1580.4%** | **+1291.5%** |
| Education | **+241.7%** | **+510.5%** | **+268.8%** |
| Sustainability | **+75.8%** | **+363.4%** | **+287.6%** |
| Natural language | **+135.4%** | **+198.7%** | **+63.3%** |
| Drug discovery | +9.7% | **+149.3%** | **+139.6%** |
| Patient | +25.5% | **+143.7%** | **+118.2%** |
| Food | **+46.4%** | **+142.6%** | **+96.2%** |
| Recommendation | **+71.1%** | **+127.6%** | **+56.5%** |
| Medical | **+163.1%** | **+115.0%** | **-48.1%** |
| Healthcare | **+291.0%** | **+57.4%** | **-233.6%** |

*Source: OpenAlex API. Sample: 4,000 China-affiliated pre-period, 5,000 post-period; global baseline 6,400 pre, 8,000 post. Computer Science identified via OpenAlex concept C41008148. Abstract coverage ≈65–70%. Term lists defined a priori. Preliminary analysis--replication at full population scale pending. Table constructed by Claude.ai*

Acknowledgement

The author used Claude (Anthropic) to assist with data retrieval scripting, bibliometric analysis, figure production, and draft editing. All analytical interpretations, theoretical arguments, and conclusions are the author's own. The author reviewed and takes full responsibility for all content.